\documentclass[conference]{IEEEtran}
\IEEEoverridecommandlockouts
\usepackage{cite}
\usepackage{amsmath,amssymb,amsfonts}
\usepackage{bbm}
\usepackage{algorithmic}
\usepackage{graphicx}
\usepackage{textcomp}
\usepackage{xcolor}
\usepackage{dsfont}
\usepackage{physics}
\usepackage{comment}
\usepackage{amsmath,amssymb,amsfonts,amsthm}
\usepackage{mathtools}
\usepackage{siunitx}
\usepackage{booktabs}

\usepackage[acronym]{glossaries}
\newacronym{awgn}{AWGN}{Additive White Gaussian Noise}
\newacronym{elroi}{ELROI}{Extremely Low-Resource Optical Identifier}
\newacronym{jdr}{JDR}{Joint Detection Receiver}
\newacronym{ssr}{SSR}{Single Shot Receiver}
\newacronym{ttr}{TTR}{Time To Read}
\newacronym{atw}{ATW}{Active Time Window}
\newacronym{leo}{LEO}{Low Earth Orbit}
\newacronym{slp}{SLP}{Satellite License Plate}

\def\BibTeX{{\rm B\kern-.05em{\sc i\kern-.025em b}\kern-.08em
    T\kern-.1667em\lower.7ex\hbox{E}\kern-.125emX}}

\begin{document}

\title{Quantum Limits of LEO Satellite Beacon Reading
\thanks{This work was financed by the DFG via grant NO 1129/2-1 and by the Federal Ministry of Research, Technology and Space of Germany via grants 16KISQ093, 16KISQ039, 16KISR026 and 16KISQ077. The generous support of the state of Bavaria via Munich Quantum Valley, the NeQuS- and the 6GQT project is greatly appreciated. Finally, the authors acknowledge the financial support by the Federal Ministry of Education and Research of Germany in the programme of “Souverän. Digital. Vernetzt.”. Joint project 6G-life, project identification number: 16KISK002. }
}
\author{
\IEEEauthorblockN{
    Pere Munar-Vallespir\IEEEauthorrefmark{1},
    Marc Geitz\IEEEauthorrefmark{2},
    \'Angeles V\'azquez-Castro\IEEEauthorrefmark{3},
    Janis Nötzel\IEEEauthorrefmark{1},
}
\IEEEauthorblockA{\IEEEauthorrefmark{1}Theoretical Quantum System Design group (TQSD), Technical University of Munich, Germany, pere.munar@tum.de}
\IEEEauthorblockA{\IEEEauthorrefmark{2}Deutsche Telekom T-Labs, Berlin, Germany}
\IEEEauthorblockA{\IEEEauthorrefmark{3}Universitat Autonoma de Barcelona, Barcelona, Spain}
}

\maketitle

\begin{abstract}
    We study the quantum limits of the \acrlong{elroi} beacon concept introduced by Holmes, Weaver, and Palmer. In this concept, a satellite continuously emits a weak optical signal to broadcast its identity. Via analysis of the fundamental limits on communication introduced by Shannon, Gordon, and Holevo, we demonstrate that in such scenarios, incorporating quantum technology into the design of a ground station significantly enhances performance. Specifically, the \acrlong{ttr} the beacon signal and thereby identify the satellite is greatly reduced in situations where weather conditions obstruct the signal, allowing the \acrlong{atw}, during which the satellite can be utilized for subsequent network operations, to be extended by nearly a factor of 20. In this particular case, the quantum technology concept that is employed is the so-called \acrlong{jdr}, which is a system aiming to operate at the Gordon-Holevo limit by performing joint quantum operations on sequences of incoming signals. 
\end{abstract}

\begin{IEEEkeywords}
Quantum Space Communication, Quantum Information, Satellite Communication
\end{IEEEkeywords}
\section{Introduction}
Satellites have become an extremely useful resource for several tasks, such as acting as data relays, observing Earth, or monitoring atmospheric activity. To communicate efficiently with a satellite, its identity and position are needed. The sudden increase in the number of satellites in the last decade, especially in \gls{leo} has made this task considerably harder. Constellations consisting of a huge number of satellites are being deployed, which cooperate to perform a single task. For example, Starlink, a service providing remote areas with Internet access, has 7578 satellites in orbit and plans to increase the number up to 42000\cite{Mohan_2024}. A constellation of up to 337.320 satellites, known as Cinnamon-937 was filed for ITU in 2021\cite{number_satellites}. LEO satellites tend to be small units with stringent restrictions on weight and power consumption. Limited by their heat dissipation and power generation, they require an efficient system to identify\footnote{Identification here refers to the physical kind, not to information theory or code-based identification.} them, given their small size, small power, fast pass times, and high number. 

The need for a system to identify satellites has led to different proposals to address this problem. One approach involves attaching low-intensity beacons to each satellite, which emit a unique, repeating pattern. This is the core idea behind the \gls{elroi} \cite{Palmer_2018,palmer2019l}. The satellite size introduces stringent power restrictions on the transmission power of the beacon. The emitted signal has a wide half-angle, as there is no mechanism to ensure that the beacon is always pointing towards the ground station. We are assuming a laser beacon, which is widened by an optical setup to create an isotropic radiation pattern. Another approach is to use passive systems on satellites, such as the \gls{slp}\cite{Bakker2024-qd}. \gls{slp} is based on reading the spectral-encoded information of a reflecting tag on the satellite by using a laser source in the ground station. While there are different ideas on how to solve this problem, both technologies have a bottleneck due to the low amount of signal light that must be read on the ground station. In the case of \gls{elroi}, due to low power on small satellites, and in the case of SLP, due to the high losses of reflection and the long optical path. 

On the other hand, quantum technologies have steadily developed over the last decade, with significant advances in quantum computing, quantum sensing, and quantum communication. A good example is the development of integrated photonics, leading to efficient and small-scale high-quality photonic components \cite{Pelucchi2022}. Since these technologies are achieving a certain degree of maturity, it is interesting to see where they could fit in the current technological landscape, in classical infrastructure. 

In Shannon theory, the \gls{awgn} channel is intended as a simple model to show how electromagnetic waves lose amplitude and get noisy as they propagate through space. Gaussian noise is the simplest model of noise and is used consistently in free-space communication. The problem of its capacity was studied by Hartley and Shannon, and it is given now as the well-known Hartley-Shannon theorem \cite{shannon}\cite{hartley}. For low received signal power, these formulas need to be modified by taking into account the shot noise arising in the detection process \cite{banaszek}, leading to the expression
\begin{equation}\label{eqn:shannon-1}
	C_{S} = \frac{1}{2} \log\left(1 + \tfrac{4 \gamma E}{2N +1}\right)
\end{equation}
for the capacity of \gls{awgn} in bits per channel use if homodyne measurement is performed at the receiver, where $E$ is the average amount of energy the sender is allowed per channel use in photons, $N$ is the amount of noise photons and $\gamma$ is the transmittance of the channel. If heterodyne measurement is used instead, then the capacity takes the form 
\begin{align}\label{eqn:shannon-2}
	C_{S'} = \log\left(1 + \tfrac{\gamma E}{N +1}\right).
\end{align}
Since $C_S>C_{S'}$ for low received power, we focus on $C_S$ in this work. We refer to the hypothetical state-of-the-art receiver performing symbol-by-symbol detection prior to (electronic) error correction and achieving rate $C_S$ as a \gls{ssr}.
While useful, this model does not represent quantum states as elements of a Hilbert space and consequently omits several key properties. The shot noise limit must be introduced artificially, by separating the noise into background noise and an additional fixed component attributed to Heisenberg uncertainty. Moreover, many important families of quantum states, such as single-photon states or squeezed states, cannot be described within this framework.

Limitations of classical models motivate the use of quantum information theory, where quantum states are represented by positive, trace-preserving operators on a Hilbert space. For optical systems, the propagating light pulses in the paraxial regime are modeled on infinite-dimensional Hilbert spaces \cite{quantum_shannon}. In this framework, laser outputs can be approximated by sequences of orthogonal optical field states. The classical capacities of optical quantum channels, also known as bosonic quantum channels, have been extensively studied, with exact results for several channel families. In general, the Holevo-Schumacher-Westmoreland theorem \cite{holevo,westmoreland} shows that capacity requires optimization over all joint inputs, including entangled ones across multiple channel uses, as quantum channels can be super-additive.

Considering photon loss during free-space propagation and the presence of thermal background light at the detector, one can model the system using the lossy bosonic channel with thermal noise - the quantum analogue of the \gls{awgn} channel \cite{giovannetti2004classical, guha2004classical}. This analogy holds in the sense that, when using coherent states as inputs (the quantum description of classical light) and choosing suitable measurements, the output statistics match those of the classical \gls{awgn} model. This channel has capacity \cite[Supplementary material]{Giovannetti_2014}
\begin{equation}\label{eqn:holevo-capacity}
	C_{H}(\gamma, E, N) = g(\gamma E + N)-g(N)
\end{equation}
in bits per channel use, where 
\begin{equation}
\label{eq : gordon}
    g(x) = (x+1)\log(x+1) - x\log(x)
\end{equation}  
is known as the Gordon function and it is also the entropy of a thermal state with mean photon number $x$. We take $x\log(x)$ to be 0 for $x=0$, as usual in information theory. $\gamma$ is the channel transmittance, $E$ the average energy that the sender is allowed to use per channel use in number of photons and $N$ is the number of noise photons per channel use. No distribution of entangled states is needed to achieve this capacity. Rather, it is known that modulation of coherent laser light is sufficient to achieve the $C_H$, just as it achieves $C_S$ \cite[Theorem 4]{rosati2017decodingprotocolsclassicalcommunication}. However, at the receiver side a collective measurement over several incoming signals must be performed to reach $C_H$, a process which is usually referred to as joint detection in the literature. Corresponding receivers are called \gls{jdr}s and have not been constructed yet, but are rather an open line of research \cite{guha, Delaney_2022, Cui2025} showing first implementations, algorithmic approaches to stabilization \cite{goekhanPhase} and new directions towards improving the spectral efficiency of current \glspl{jdr} \cite{Rosati:24}.

The capacity of the classical model and the quantum model becomes asymptotically close for large signal energies or a large amount of noise, which can be seen by taking the respective limits. However, the situation is different at low photon regimes. If the amount of photons per incoming signal is close to $E=1$ or smaller, then quantum effects start to become relevant. A \gls{jdr} can, in principle, achieve a better performance in parameter regions where signal energy is low but background noise is also low. 

In addition to specific ``indoor wireless internet of things'' and ``underwater internet of things'' type of applications as described in \cite{amiri2024}, we thus identify a third scenario of potential application in beacon reading.

\section{Problem definition}
Assuming a number $S$ of satellites in a constellation, we consider the problem of reading the beacon ID of a given satellite by detecting and analyzing its (possibly weak) electromagnetic beacon signal with a receiver located in a base station. In the terminology of information theory, this is equivalent to distinguishing a number $S$ of signals reliably. 
We compare two hypothetical receivers: One conventional system with the ability to operate at the Shannon limit and one based on quantum technology, which can operate at the Holevo limit. The first one is called \gls{ssr}, the second one is called a \gls{jdr}. The system's key parameters are based on the work \cite{Palmer_2018} and are listed in Table \ref{tab:parameters}. We further assume the size of our constellation is set to $S=10^6$. This implies that $20$ bits are needed to enumerate the satellites, and every satellite thereby has associated with it a unique binary string of length $20$, which we refer to as beacon ID.
We assume each satellite system uses a fixed, error-corrected sequence that it transmits repeatedly. We consider only loss and non-coherent background radiation as a source of error. In order for the base station to be able to read the sequence, the sequence is designed to be decodable under an assumed worst-case loss using a compound code \cite{seitzCompound}. If this code has rate $R$ bits per second, then the time it takes the satellite to transmit the $20$ bits of its identifier is \gls{ttr}$=20/R$. After the satellite has been identified, the base station proceeds to using it for other network tasks, until it vanishes again. We call this time the \gls{atw}. Assume a satellite rises at the horizon from the perspective of the base station at time $t_1$ at a distance $d_1$, and passes by directly over the base station at time $t_2>t_1$ at a distance $d_2<d_1$. Assuming the satellite takes a time $T_s$ to rise and vanish again, the \gls{atw} is therefore defined as \gls{atw}$:=T_s-t-1.5\cdot$\gls{ttr}, where the factor $1.5$ accounts for the fact that, on average, the base station starts reading the transmitted sequence halfway into that sequence, and $t$ is the time where the base station starts reading the sequence. Under fixed ideal weather conditions (no loss from atmospheric conditions), the following three situations can occur: 

\begin{enumerate}
    \item The code is designed to work for the loss corresponding to link distance $d_2$. Then the \gls{atw} is upper bounded by half the passing time. In addition, even small deviations lead to complete inability to read the beacon, resulting in \gls{atw} being equal to $0$.
    \item The code is designed to work for the loss corresponding to distance $d_1$. In this case, it can happen that the \gls{ttr} exceeds the passing time, resulting again in an \gls{atw} equal to 0.
    \item The code is adapted to a link loss corresponding to a distance $d$ satisfying $d_1<d<d_2$. In this case, the optimal \gls{atw} may be reached. 
\end{enumerate}

In this scenario, a system using a \gls{jdr} for the base station design will typically have a lower \gls{ttr} than a conventional one. We analyze the impact of the design choice between the \gls{jdr} and a conventional receiver on the \gls{atw} . 
\section{Analysis}
\paragraph{Impact of second-order corrections} The fundamental limit on the number of bits that can be transmitted over a channel with capacity $C$ with vanishing error in $n$ uses is asymptotically given by $n\cdot C$. The real-world impact of using short codes is quantified by finite block-length corrections, which relate the decoding error, code length, and number of different messages. The respective bounds for conventional systems are well known from \cite{Polyanskiy2010}, and shot noise can be accounted for as in \cite{amiriTrading}. For the quantum second-order coding rate, a lower bound is known in the situation where no thermal noise affects the system \cite{Wilde_2015}. In the situation investigated in this work, the number of channel uses is in the order of $n=10^6$, and second-order effects were numerically found to play no role in the analysis. Thus, we focus solely on capacities.
\paragraph{System design}
Besides those outlined in \cite{Palmer_2018}, several other factors are relevant. One is the design complexity of the link, which is comparable on the transmitter side for both the conventional system and the one employing a \gls{jdr}. However, on the receiver side, the \gls{jdr} introduces added complexity due to the need for specific quantum operations, such as non-destructive photon counting \cite{giovannetti2012} or other specifically tailored light-matter interactions \cite{polarCodes}. In the present scenario, this is immediately obvious via the requirement for extremely narrow-band filters. These filters are crucial for minimizing the number 
$N$ of noise photons, as the theoretically infinite advantage of the \gls{jdr}  - derived from equations \eqref{eqn:holevo-capacity} and \eqref{eqn:shannon-2} - only emerges in scenarios with both low received signal power and minimal noise.
\subsection{Base Scenario}
We consider the system as described in \cite{Palmer_2018}. The time it takes a satellite orbiting Earth at $1.000$km distance to rise from the horizon, pass by directly on top of the ground station, and vanish again behind the horizon is $T=1054$ seconds. We note that this is an idealistic assumption which underestimates the value of the \gls{jdr}: in fact, the \gls{atw} can be significantly lower in cases where the satellite trajectory does not pass by exactly on top of the base station (resulting in a lower value for $T$). For example, at 400km orbital altitude and 45 degrees above the horizon, where some satellites used for remote sensing are situated, $T$ is already reduced to 104s\cite[Table 1]{leopasstimes}.

A critical value is the link loss, which is quantified by the number $\gamma$ and, in the given context, computed as 
\begin{align}
	\gamma = \frac{A\cdot\tau\cdot \epsilon_{DQE}}{\Omega\cdot r^2}
\end{align}
where $A$ is the receiver aperture, $\tau$ the combined loss of atmosphere and filter, $\epsilon_{DQE}$ the quantum efficiency, $\Omega$ the solid angle of the beacon emission and $r$ the distance between satellite and ground station.

The number $N$ can, in the given context, be computed from \cite{Palmer_2018}, where $91$ noise photons per second have been reported at a filter width of 10nm. We assume this filter to enable transmission in the range from 637.5nm to 638.5nm, which corresponds to frequencies from 470.262,7GHz down to 469.526,2GHz. For a pulse duration of 2$\mu$s, however, less than 1GHz bandwidth is needed. Assuming future technology, the filter bandwidth could therefore be reduced by $6$ orders of magnitude. The system design in \cite{Palmer_2018} is made using the parameters of Table \ref{tab:parameters}. 
\begin{table}[h!]
	\centering
	\begin{tabular}{ll}
		\toprule
		\textbf{Link budget parameter} & \textbf{Value} \\
		\midrule
		Beacon wavelength & $\lambda$ = \SI{638}{\nano\meter} \\
		Peak power emitted & $P_{\text{peak}}$ = \SI{1}{\watt} \\
		Pulse width & $\tau$ = \SI{2}{\micro\second} \\
		Pulse interval & $T$ = \SI{500}{\micro\second} \\
		Fraction of 1 bits & $f_1$ = 0.50 \\
		Solid angle of emission & $\Omega$ = $2\pi$ \\
		Distance to receiver & $r$ = \SI{1000}{\kilo\meter} \\
		Diameter of receiver telescope & $D$ = \SI{36}{\centi\meter} \\
		Filter transmission at $\lambda$ & $T_f$ = 0.83 \\
		Filter bandwidth & $\Delta \lambda$ = \SI{10}{\nano\meter} \\
		Solar spectral flux at $\lambda$ & $I_\lambda$ = \SI{1.654}{\watt\per\meter\squared\per\nano\meter} \\
		Detector quantum efficiency & $\varepsilon_{\text{DQE}}$ = 0.039 \\
		10-cm Cubesat effective albedo area & $\alpha_{\text{CS}}$ = \SI{0.00053}{\meter\squared} \\
		1-m satellite effective albedo area & $\alpha_{\text{1m}}$ = \SI{0.053}{\meter\squared} \\
		\bottomrule
		\end{tabular}
        \hspace{8mm}
	\caption{\label{tab:parameters}Link budget parameters from \cite{Palmer_2018}}
\end{table}
From these numbers, by choosing a less ambitious filter width, we use instead $\Delta \lambda'=10^{-4}$nm. Assuming zero dark-counts, we then arrive at a modified number of $0.01$ noise photons per second while retaining the original number of $3$ signal photons per second \cite[Table II]{Palmer_2018}. By introducing the modulation speed (number of symbols per second) B, we can transform capacities from bits per channel use to bits per second via $C(\gamma,E,N)\to B\cdot C(\gamma,E/B,N/B)$. At a modulation speed of B=1MHz and a transmittivity of $\gamma=5.15\cdot10^{-15}$ \cite{Palmer_2018}, this amounts to a Holevo (Shannon) capacity of 
\begin{align}
	10^{6}\cdot C_{H}(\gamma,6.4\cdot10^{9},10^{-3})\approx59.27\\
	10^{6}\cdot C_{S}(\gamma,6.4\cdot10^{9},10^{-3})\approx8.65
\end{align}
bits per \emph{second}, so that an almost seven-fold capacity advantage can be realized for such a system. Note that due to the low received photon number, the respective  capacities are extremely low if compared with the Giga- or Terahertz capacities of standard optical communication links today. As a design goal, we require our system to be able to reliably distinguish one million satellites. To enumerate these $10^6$ satellites, $20$ bits are needed. We can therefore calculate the number of seconds that are needed to transmit these $20$ bits, which we take as the \gls{ttr}. We conclude that the quantum system takes only $\approx0.34$ seconds to read the beacon ID, while the conventional system needs around $2.31$ seconds if operating at the Shannon limit. A comparison with the reading time of up to $150$s reported in \cite{Palmer_2018} indicates that \gls{ttr} may be much longer in practice. 
We now proceed to a systematic study of the impact of the use of the \gls{jdr} on the \gls{ttr} and \gls{atw}.

\subsection{Availability}
High availability of a satellite and little delay in reading its beacon ID are desirable features. Therefore, we consider the \gls{ttr}, the time it takes the ground station to read the beacon ID, to be an important parameter in the system design. 
\begin{figure}[ht]
	\includegraphics[width=.5\textwidth]{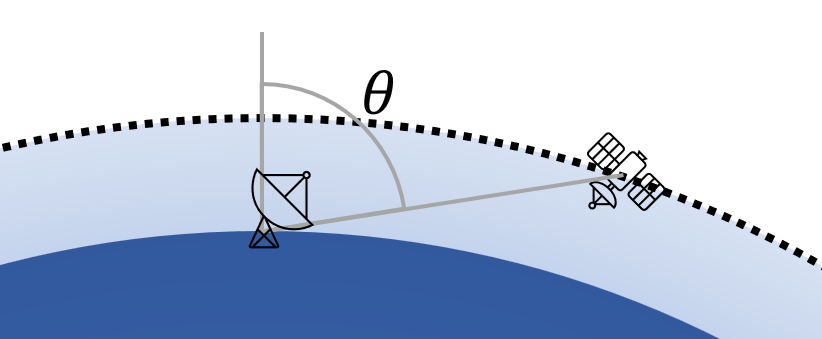}
	\caption{\label{fig:satellite-system}\gls{leo} satellite and base station. Depicted is the elevation angle $\theta$.}
\end{figure}
In the situation of the \gls{elroi} system (as depicted in Figure \ref{fig:satellite-system}), the initial design goal \cite{Palmer_2018} is to read the beacon ID within one passage of the satellite, for which a time of $1054$s can be computed as the maximum possible time window given an orbital height of $1.000$km. The reported times of up to $157s$ \cite{Palmer_2018} are sufficient to identify the satellite while it passes by, but are a significant fraction of the time the satellite is visible. 

Assuming the \gls{elroi} system is not adapting to particular distances to particular ground stations or to specific weather conditions, the design should be made for a specified worst-case scenario, following the concept of compound channels\cite{seitzCompound}. In compound channels, the worst-case scenario for the channel conditions dictates the achievable rate. This implies that the worst-case link budget for which the beacon ID can still be read determines the \gls{ttr}. Note that a satellite at an orbital height of $1,000$km is $2,842$km away from the base station when it is at $4.5$° (0.1 radians) over the horizon. In Table \ref{tab:system-performance}, we list the classical (C) and quantum (Q) channel capacity in bits per second and the \gls{ttr} in seconds, respectively, for several scenarios: We vary the distance between $1,000$ and $2,842$ km and weather conditions between transmission through vacuum, through atmosphere, and through light rain. Transmission through the atmosphere and light rain adds extra attenuation. Absorption in rainy conditions has been reported to reach as high as 1dB/m at 650nm in \cite{Li:18}.
 Since no measurements are available exactly at 638nm, we instead use the values reported for 650nm. Given the reported value of 1dB/m, assuming a total loss of $22$dB resulting from slightly adverse atmospheric conditions seems to be a conservative estimate. 
The attenuation caused by the atmosphere itself has been reported to be approximately 0.5dB/km \cite{athmosphericLoss}, so an additional loss of 10dB can as well be considered a conservative estimate.

\begin{figure}[ht]
	\includegraphics[width=.5\textwidth]{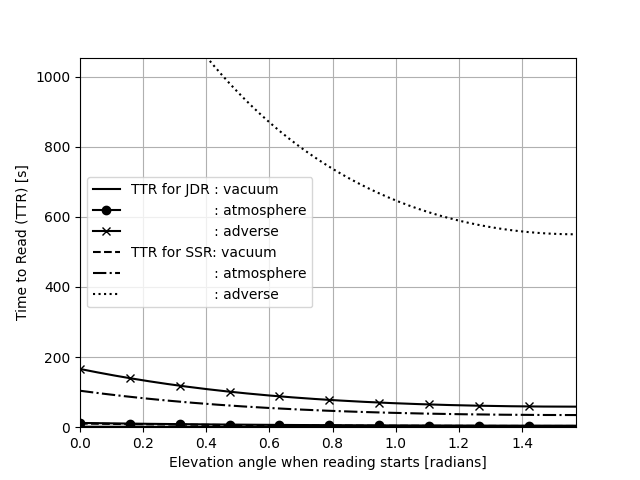}
	\caption{\label{fig:radians-to-ttr} Shown is the time in seconds that the ground station requires to read the beacon ID, referred to as the \gls{ttr}, plotted against the satellite's elevation angle above the horizon at the moment the reading begins when the satellite's orbital height is $1,000$ km.}
\end{figure}

\begin{figure}[ht]
    \includegraphics[width=.5\textwidth]{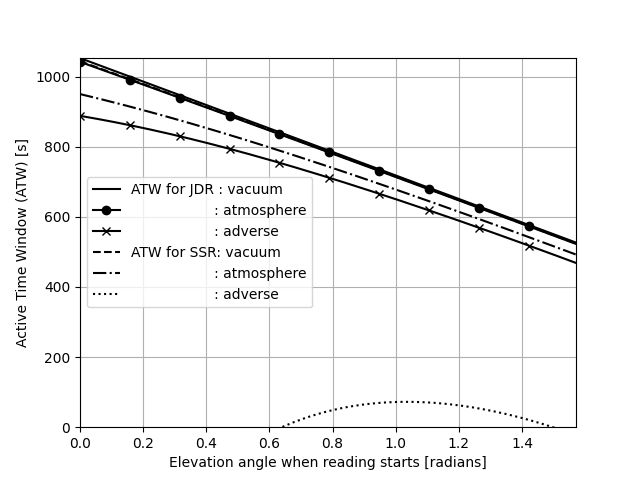}
    \caption{\label{fig:radians-to-atw}
        The figure shows the \gls{atw} of the satellite as a function of the elevation angle at which the ground station begins to read the beacon ID when the satellite's orbital height is $1,000$ km. 
    }
\end{figure}

In Fig. \ref{fig:radians-to-ttr} we show the \gls{ttr} as a function of elevation angle at which the base station starts to read the signal when no atmospheric or weather conditions are accounted for. An initial offset of the receiver is accounted for by applying a factor of 1.5 to \gls{ttr}. Solid black lines represent the \gls{ttr} with quantum technology, while dashed black lines represent it without. $T_s$ is set to 1045 seconds (corresponding to the highest displayed value on the y axis), corresponding to the time it takes the satellite to move from 0 to $\pi$ radians in elevation angle. For example, the dotted line represents the \gls{ttr} of an \gls{ssr}, which will not be able to decode the beacon of the satellite by any means. In other weather conditions, the \gls{jdr} gives a \gls{ttr} advantage, as depicted in Table \ref{tab:system-performance}. In Fig. \ref{fig:radians-to-atw} we display the corresponding \gls{atw}. Notably, the \gls{atw} reaches nearly 100\% of the total available time when the \gls{jdr} is employed, whereas the conventional system exhibits significantly reduced \gls{atw} under adverse weather conditions. In a cloudy environment using the \gls{ssr}, the \gls{atw} drops to zero if the code is optimized for low elevation angles. In such cases, the time required for decoding, including error correction, exceeds $T_s$, rendering successful beacon message decoding practically impossible. Contrary to common expectations, quantum communication proves to be the more robust approach under these challenging conditions.

\begin{table}[ht]
\centering
\scriptsize
\caption{Performance Comparison: Classical vs Quantum Systems}
\label{tab:system-performance}
\begin{tabular}{@{}lccccc@{}}
\toprule
Scenario & Dist. & Extra & Capacity (C/Q) & \gls{ttr}(C/Q) \\
         & (km)  & (dB)  & (bit/s)    & (s)      \\
\midrule
Zenith     & 1,000 & 0    & 8.65 / 59.27   & 2.31 / 0.34   \\
Early     & 2,000 & 0    & 2.16 / 16.33   & 10.39 /1.40  \\
Horizon    & 2,842 & 0 & 1.07 / 8.46 & 21.96 / 2.97 \\
Atmosphere    & 1,000 & 10 & 0.22 / 1.87    & 103.97 / 12.15    \\ 
Adverse weather    & 1,000 & 22 & 0.01 / 0.13    & 1650.35/ 166.07   \\ 
\bottomrule
\end{tabular}

\end{table}

\subsection{Network Aspects}
    For example, a constellation of $ 10^6$ satellites at an orbital altitude of 1,000 km, coarse-grained estimates—assuming that each satellite remains within the base station’s field of view for 1,054 seconds—indicate that more than 100 satellites enter this area every second. If the base station must identify multiple satellites before selecting one capable of performing a given task, these estimates highlight the critical importance of short \glspl{ttr}. For a satellite emerging at the horizon, the \gls{ttr} of the \gls{jdr} is approximately 3 seconds, whereas that of the \gls{ssr} is around 22 seconds (see Table \ref{tab:system-performance}).

\subsection{Design Aspects and Open Problems}
\paragraph{Frequency Filters}
In the conventional system, with the number of noise photons below 90 per second and a modulation speed of 1MHz, the resulting signal-to-noise ratio (SNR) $2\gamma E/(4N+B)$ (see equation \eqref{eqn:shannon-1}) is only marginally influenced by the actual photon number - reducing the filter bandwidth has little to no effect.
To the contrary, the performance of a \gls{jdr} is highly sensitive to background noise. In our base scenario ``Zenith'' in Tab. \ref{tab:system-performance}, the Holevo capacity drops to around 40 bits per second when the background noise photon number per second is set to N=90. Thus, multiple system components need to be jointly optimized to support advanced quantum functionality.
\paragraph{Quantum Technology}
A version of the \gls{jdr} which is based on short-term quantum memory and non-destructive photon counting \cite{giovannetti2012} is believed to be inefficient due to its high complexity: This system loops through all possible messages to identify the right one, leading to exponential complexity of the decoder in the number of gates. Depending on the signal bandwidth, the approach may, however, be suitable from the perspective of decoding time -- if the signal quality can be kept high during the decoding process. Thus, the short-term quantum memory quality can become a crucial part of the design. Several physical phenomena are already being exploited in the development of quantum memories for optical quantum states. Notable progress has been made in improving fidelities, efficiencies, storage times, and multiplexing capabilities \cite{review_quantum_memories}. Most of this progress has focused on single-photon memories. Experimental demonstrations of coherent light storage using atomic frequency combs (AFCs), which enable the storage of weak coherent pulses—similar to the signals expected in our scenario, have been reported recently \cite{Duranti_2024, aweak_pulse_storage}.
\paragraph{Network Perspective}
Since ground stations are not subject to the stringent size and energy constraints faced by satellites, they are well-suited to operate a complex system such as a \gls{jdr}. In situations of high loss, the capacity of the link can then be significantly increased by enhancing the receiver's complexity, while requiring only minimal changes at the transmitter. Note that the \gls{elroi} system was originally designed using non-coherent laser LEDs, chosen for their low cost and low power consumption. However, extensive research has led to the development of smaller, more compact, and more efficient lasers that could potentially be integrated into small satellites. For example, Titanium:sapphire lasers \cite{Yang2024} and erbium-based lasers \cite{liu2023fullyhybridintegratederbiumbased} have been realized using integrated optics. While their output power remains relatively low, it is already on the order of milliwatts, making them suitable for \gls{leo} applications.

\subsection{Conclusion}
We have presented a novel use case for Joint Detection Receiver (JDR) technology, specifically targeting scenarios where low-power beacons are employed to transmit control information over long distances or under adverse weather conditions. The analysis highlights the following key advantages:

\begin{enumerate}
\item Significant reduction in the \gls{ttr} for satellite beacon systems—particularly under poor channel conditions—when comparing \gls{ssr} receivers to \gls{jdr}-based receivers.
\item The ability to initiate beacon reading well before a conventional communication link becomes operational.
\item Sufficient performance even in the context of the largest planned satellite constellations.
\item Maximized utilization of the available time window for classical communication, eliminating the need to allocate valuable payload bandwidth for satellite identification.
\end{enumerate}

JDR technology enables a fundamental shift in system complexity from the transmitter to the receiver. We have emphasized the importance of narrow-band optical filters, short-term quantum memories, and non-destructive measurement techniques for the practical realization of such systems. While these enabling technologies are not yet fully mature, ongoing research shows promising progress. For instance, coherent photonic sources suitable for satellite integration are already being developed. We are optimistic that similar advancements will follow for the remaining technological challenges.

Finally, we note that our analysis is based on simplified link budget estimates and channel models. We do not treat phase noise, which considerably complicates theoretical analysis. Further refinement and experimental validation will be necessary in future work to confirm the feasibility and performance under realistic operating conditions.

\bibliographystyle{ieeetr}
\bibliography{bibtex}

\end{document}